\documentclass{article}
\usepackage[fleqn]{nccmath} 
\usepackage{spconf,amsmath,graphicx}
\usepackage{subcaption}
\usepackage{blindtext}
\usepackage{multirow}
\usepackage{comment}
\usepackage{caption}
\usepackage{enumitem}

\title{A context-aware computational approach for measuring vocal entrainment in dyadic conversations}
\name{Rimita Lahiri$^1$, Md Nasir$^2$, Catherine Lord$^3$, So Hyun Kim$^4$, Shrikanth Narayanan$^1$}
\address{
  $^1$Signal Analysis and Interpretation Laboratory, University of Southern California, Los Angeles, USA\\
  $^2$Microsoft AI for Good Research Lab, Redmond, Washington, USA\\
   $^3$Semel Institute of Neuroscience and Human Behavior, University of California, Los Angeles, USA\\
    $^4$ School of Psychology, Korea University, Seoul, Korea\\
  }

\begin{document}
%
\maketitle
\begin{abstract}
Vocal entrainment is a social adaptation mechanism in human interaction, knowledge of which can offer useful insights to an individual's cognitive-behavioral characteristics. 
We propose a context-aware approach for measuring vocal entrainment in dyadic conversations. We use conformers~(a combination of convolutional network and transformer) for capturing both short-term and long-term conversational context to model entrainment patterns in interactions across different domains. Specifically we use cross-subject attention layers to learn intra- as well as interpersonal signals from dyadic conversations. We first validate the proposed method based on  classification experiments to distinguish between \textit{real}~(consistent) and \textit{fake}~(inconsistent/shuffled) conversations.  Experimental results on interactions involving individuals with Autism Spectrum Disorder also show evidence of a statistically-significant association between the introduced entrainment measure and clinical scores relevant to symptoms, including across gender and age groups.


\end{abstract}

\vspace{-3.5ex}

\section{Introduction}
\label{sec:intro}

\vspace{-1.5ex}

Interpersonal human interactions, notably dyadic interactions~(interactions involving two people), are widely studied by social science and human-centered computing researchers alike~\cite{vinciarelli2008social,narayanan2013behavioral}. Such interactions are characterized by rich information exchange across multiple modalities including speech, language, and visual cues. Over the years, a significant amount of effort has been invested in developing tools for both conversational data collection and in understanding and modeling the signals extracted from these interactions. 


\par A phenomenon called {\em entrainment}~\cite{brennan1996lexical,levitan2011measuring} has been described as one of the major driving forces of an interaction \cite{giles1997accommodation}. While entrainment can be exhibited within and across different modalities, vocal entrainment~\cite{nasir2020modeling} or acoustic-prosodic entrainment~\cite{Lee2010QuantificationofProsodicEntrainment,levitan2011measuring,Xiao2015AnalyzingSpeechRateEntrainment} is defined as an interlocutor's tendency to accommodate or adapt to the vocal patterns of the other interlocutor over the course of the interaction. 
Understanding entrainment~\cite{lahiri2022interpersonal} can provide meaningful insights to analyze behavioral characteristics of the individual interlocutors and the interaction participants. For example, a higher degree of entrainment is associated with positive behavioral markers like social desirability, smoother interactions, higher rapport content \textit{etc.}\cite{natale1975convergence,lubold2014acoustic}.
Entrainment can also serve as a valuable instrument to characterize behaviors in the study and practice of psychiatry and developmental studies involving distressed couples, children with autism spectrum disorder, addiction, \textit{etc}~\cite{nasir2020modeling,lahiri2022interpersonal}.

Due to the complex nature of entrainment and a scarcity of appropriately labeled speech corpora, quantifying entrainment is a challenging task. Most of the early works have relied on empirical and knowledge-driven tools like correlation, recurrence analysis, time-series analysis, spectral methods to measure how much a speaker is entraining to the other speaker~\cite{delaherche2012interpersonal}. This body of work often relied on the assumption of a linear relationship between the extracted entrainment representations and vocal features, which may not always hold. 
On the other hand, although context during a conversation  plays an important part in interpersonal interactions, it has not been incorporated in existing approaches for measuring entrainment. 
While the recent line of works~\cite{nasir2020modeling} employ a more direct data-driven strategy to extract entrainment related information from raw speech features, 
such are formulated in a way that they inherently only consider short-term context while overlooking more long-term context.
Recently context-aware deep learning architectures such as transformers~\cite{vaswani2017attention} have been proposed to capture richer contexts by explicitly modeling the temporal dimension and found many applications in natural language processing, speech and vision. 
In light of their success in modeling rich temporal context, we investigate if transformers can help capture meaningful information for quantifying entrainment.


\par In this work, we develop a context-aware model for computing entrainment, addressing the need for both short and long-range temporal context modeling. For the scope of this work, the proposed framework incorporates `context' by aiming to train the model to learn the influence of the speakers on each other. We follow the established strategy of using a distance-based measure between consecutive turn-pairs in the projected embedding space and introduce the \textit{Contextual Entrainment Distance~(CED)} metric. The main contributions of this work are two fold: first, we use a combination of self-attention and convolution to extract both short-term and long-term contextual information related to entrainment; and second, we propose a transformer-based cross-subject framework for joint modeling of the interacting speakers to learn the pattern of entrainment. We experimentally evaluate the validity and efficacy of CED in dyadic conversations involving children and study its association with respect to different clinically-significant behavioral ratings where the role of entrainment has been previously implicated~\cite{lahiri2022interpersonal}. 

\begin{figure}
    \centering
    \includegraphics[width=0.5\textwidth]{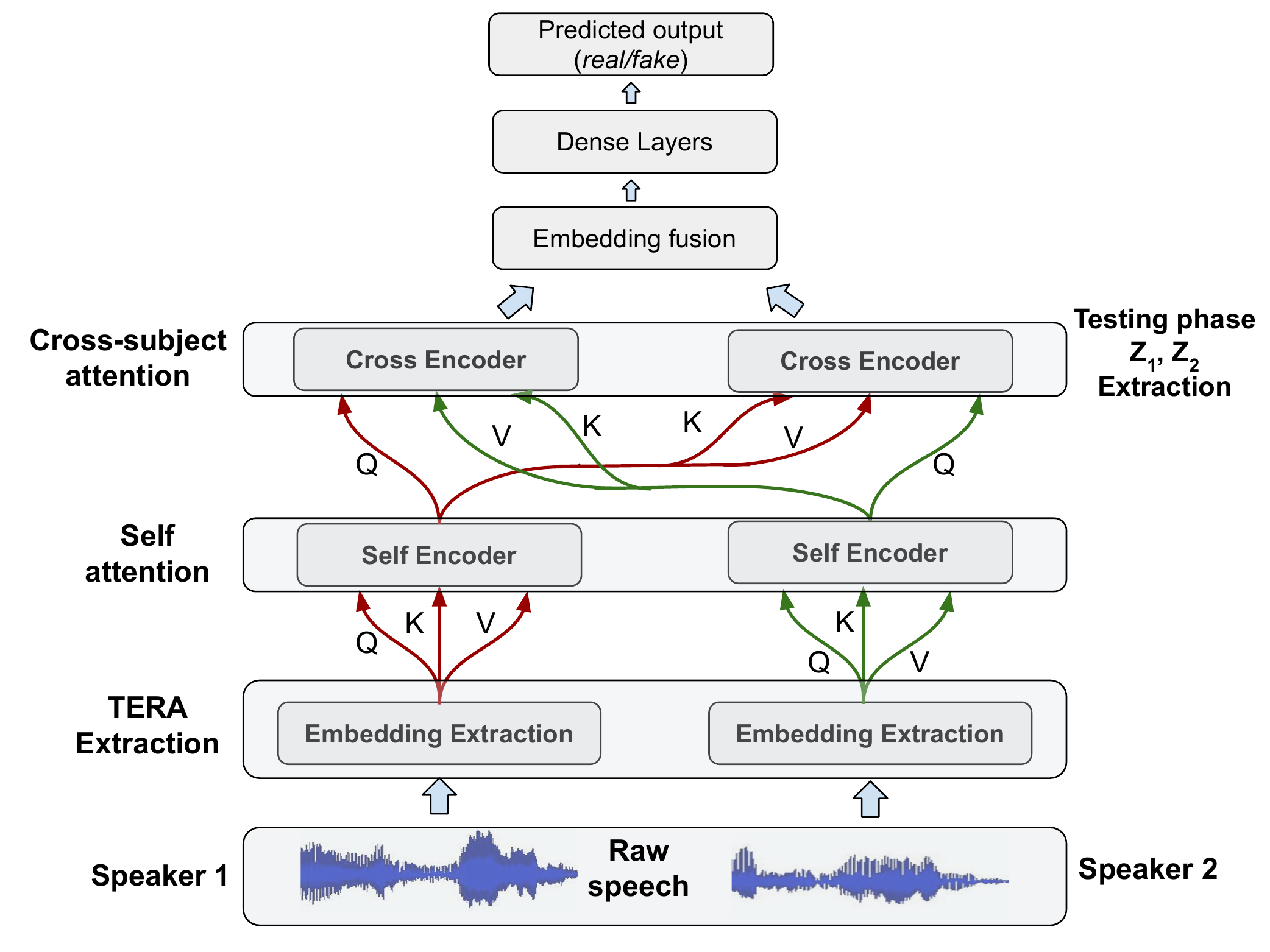}
    \vspace{-3.5ex}
    \caption{Architecture for CED extraction.}
    \label{fig:model}
\vspace{-3.5ex}
\end{figure}

\vspace{-2.5ex}

\section{Computing context-aware entrainment measure}
\label{sec:method}


\subsection{Unsupervised model training and CED computation}

Prior literature in this domain have relied on computing a distance measure directly between the turn-level speech features $X_1$ and $X_2$ from speaker~$1$ and speaker~$2$ respectively~\cite{levitan2011measuring}. However, these features also capture additional information such as speaker characteristics and ambient acoustic information which do not contribute towards learning the target entrainment patterns. The objective is to learn the inverse mapping between the embedding space~($Z_1$,$Z_2$) and the feature space~($X_1$,$X_2$) such that the model should learn to recognize turn pairs with high and low levels of entrainment.

Here, we formulate the problem by training the network to classify between interactions having consecutive turn segments~\textit{(true samples)} and interactions having random/shuffled turn segments~\textit{(fake samples)}. We temporally partition the conversational audio sequence into speaker specific chunks and feed these chunks to the model to predict whether the fed audio chunks are part of real conversation or a fake one. 

After the training phase, we use the trained network weights to extract the cross-encoder layer outputs for both speakers. Next, we calculate CED as the smooth L1 distance~\cite{nasir2020modeling} between the embeddings obtained in the previous step.

\vspace{-2.5ex}
\subsection{Model architecture}

As shown in Fig. \ref{fig:model}, we use two main modules to build the model to compute entrainment, first, the self-attention encoder that is used to enhance the extracted features by attending to themselves and then, a cross-attention encoder which allows the features to attend to a different source.

We use conformer~\cite{gulati2020conformer} layers for the self-attention module to model both short-term and long-term dependencies within an audio sequence in a parameter-efficient way by incorporating a convolutional module in the transformer layer. The self-attention layer obtains meaningful representation from the long-term interaction and the convolution layer is used to learn the local relation amongst the interaction based features.

To extract meaningful information related to entrainment, previous works have mostly relied on individual modeling of interlocutors involved in a conversation. However, entrainment being an interpersonal phenomenon, the need for jointly modeling interlocutors becomes heightened in such scenarios. We address this issue by using a transformer layer for cross-subject attention, allowing the features extracted per subject to access each other to capture crossed influence over the interaction.

\vspace{-2.5ex}
\section{Experiments}
\label{sec:experiment}

\vspace{-2.5ex}

\subsection{Datasets}


We use the following two datasets for our experiments.

\textit{The Fisher Corpus} English Part 1 (LDC2004S13)~\cite{cieri2004fisher} consists of spontaneous telephonic conversations between two native English speaking subjects. There are 5850 conversations of approximately 10 minutes duration. The dataset is accompanied with transcripts along with timestamps marking speaker duration boundaries. We use 60\% of this dataset for training and 5\% for testing.

\textit{The ADOSmod3 corpus} consists of recorded conversations from autism diagnostic sessions between a child and a clinician who is trained to observe the behavioral traits of the child related to \textit{Autism Spectrum Disorder~(ASD)}. A typical interactive session following the \textit{Autism Diagnostic Observation Schedule~(ADOS)-2} instrument lasts about 40-60 minutes, and these sessions are composed of a variety of subtasks to evoke spontaneous response from the children under different social and communicative circumstances. In this work, we consider the administration of Module 3 meant for verbally fluent children and adolescents. Moreover, we focus on \textit{Emotions} and \textit{Social difficulties and annoyance} subtasks as these are expected to extract significant spontaneous speech and reaction from the child while answering questions about different emotions and social difficulties. The corpus consists of recordings from 165 children collected across 2 different clinical sites. We use this corpus for evaluation purpose, the demographic details of the dataset are reported in Table \ref{tab:dataset}.

\begin{table}[t!]
\begin{center}
\caption{Demographic details of ADOSMod3 dataset}
\label{tab:dataset}
\resizebox{0.48\textwidth}{!}{
\begin{tabular}{c c} \hline \hline
\textbf{Category} & \textbf{Statistics}  \\  \hline
 Age(years) & Range: 3.58-13.17 (mean,std):(8.61,2.49)  \\  \hline
 Gender & 123 male, 42 female \\ \hline
 Non-verbal IQ & Range: 47-141 (mean,std):(96.01,18.79) \\  \hline
  \multirow{4}{4em}{\centering Clinical Diagnosis} & 86 ASD,42 ADHD\\  
 & 14 mood/anxiety disorder\\
 & 12 language disorder\\
 & 10 intellectual disability, 1 no diagnosis\\  \hline
 \multirow{2}{4em}{\centering Age distribution} & Cincinnati: $\leq$5yrs 7, 5-10 yrs 52, $\geq$10yrs 25 \\
 & Michigan: $\leq$5yrs 11, 5-10 yrs 42, $\geq$10yrs 28\\ \hline \hline
\end{tabular}}
\end{center}
\vspace{-3ex}
\end{table}

\vspace{-2.5ex}

\subsection{Experimental setup}

\subsubsection{Feature extraction}

 In this work, to compute CED the speech segments of interest are conversational turns from both speakers. 
 We compute the speaker turn boundaries from the time information available in the transcripts, excluding the intra-turn pauses to avoid including noisy and redundant signals. For every speaker turn,
 we extract self-supervised TERA embeddings~\cite{liu2021tera} to obtain a $768$ dimensional feature vector. We choose TERA embeddings
 as it employs a combination of auxiliary tasks to learn the embedding instead of relying on a single task, so it is expected to learn enhanced features from raw speech signals.

\begin{figure}[!tbp]
  \begin{subfigure}[b]{0.23\textwidth}
    \includegraphics[width=\textwidth]{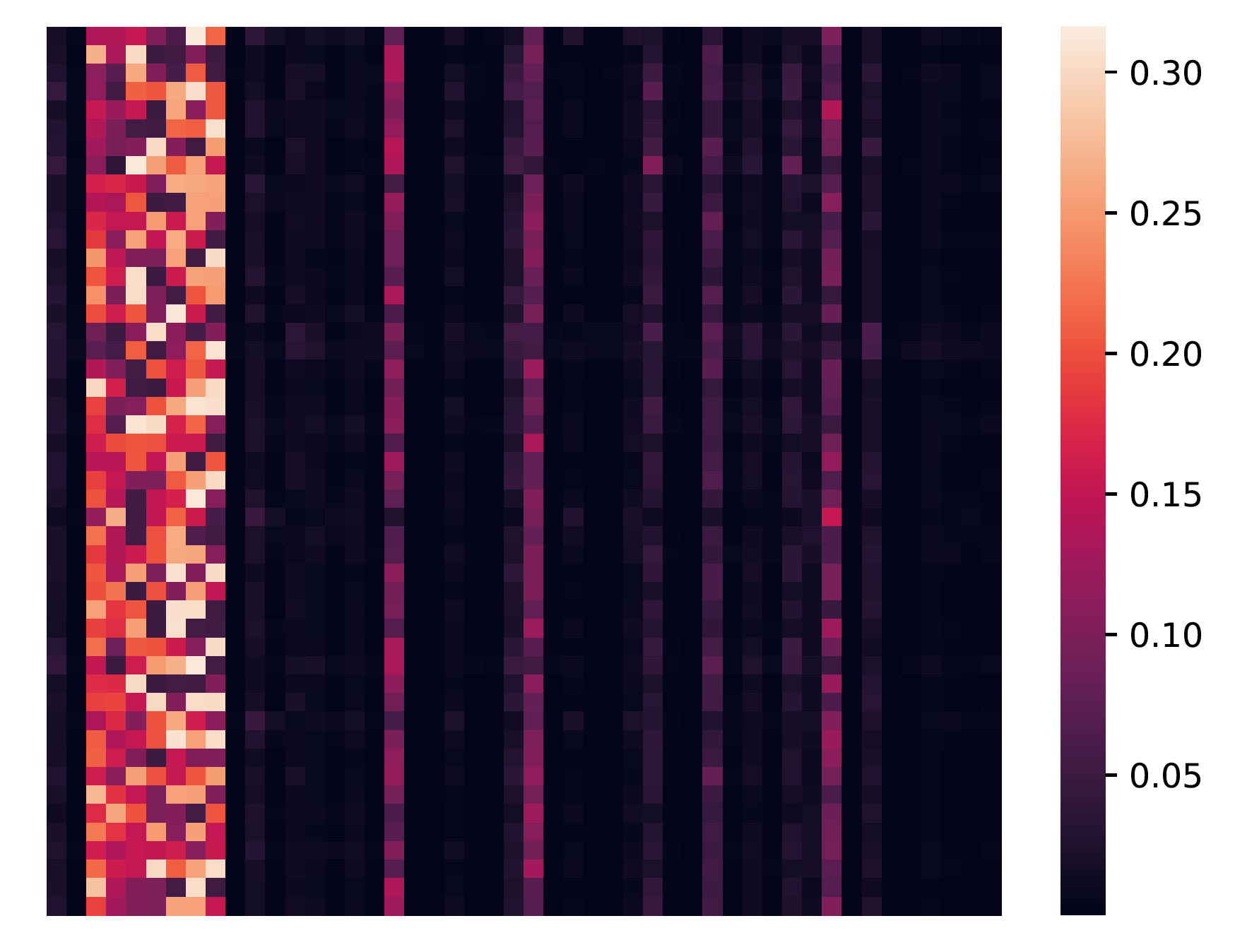}
    \caption{Cross-encoder 1}
    \label{fig:enc1}
  \end{subfigure}
  \hfill
  \begin{subfigure}[b]{0.23\textwidth}
    \includegraphics[width=\textwidth]{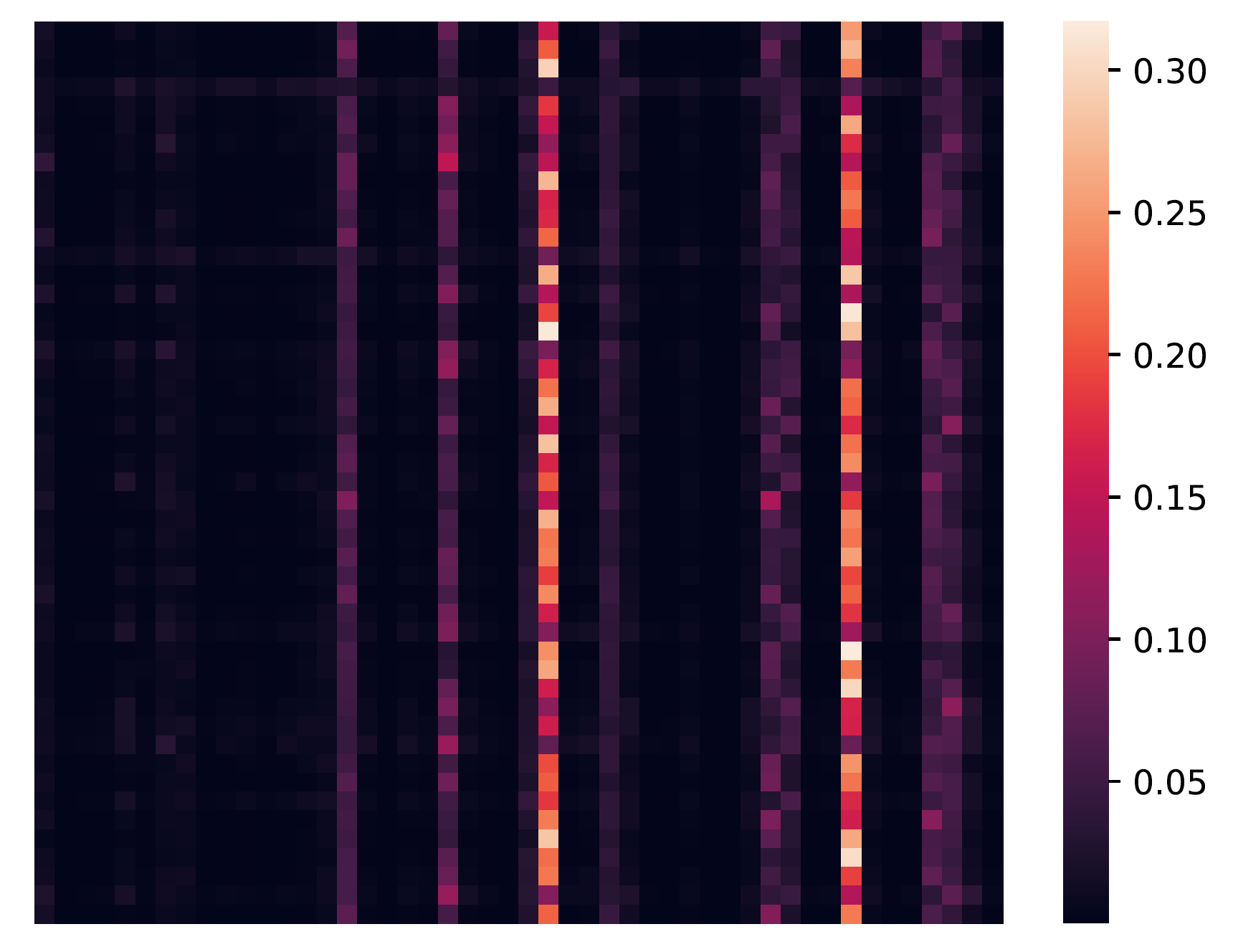}
    \caption{Cross-encoder 2}
    \label{fig:enc2}
  \end{subfigure}
  \vspace{-1ex}
  \caption{Attention activations}
  \label{fig:attention}
\end{figure}

\begin{table}[t!]
\begin{center}
\caption {Classification experiment for \textit{real vs fake} sessions}
\label{tab:classification}
\begin{tabular}{ c|c|c } 
\hline
\hline
\multirow{2}{*}{Measure} & \multicolumn{2}{|c}{Classification accuracy(\%)} \\ 
\cline{2-3}
& Fisher Corpus & ADOSMod3 Corpus \\ 
\hline
Baseline 1 & 80.52 & 82.22 \\
Baseline 2 & 76.33 & 70.64 \\ 
Baseline 3 & 82.91 & 85.73 \\ 
\hline
CED & 92.13 & 95.66 \\ 
\hline\hline
\end{tabular}
\end{center}
\vspace{-4.5ex}
\end{table}

\vspace{-2.5ex}
\subsubsection{Parameters and implementation details}

We use \textit{352} and \textit{64} attention units for the conformer and transformer layers, respectively, while \textit{4} attention heads are employed for both.
The full architecture obtained by using a conformer layer followed by a transformer layer results into \textit{2.1M} parameters.
The model is trained with a binary cross entropy with logits loss function and  Adam optimizer with the initial learning rate of $1e^{-5}$.
There is a provision of early stopping after 10 epochs if no improvement is seen in validation loss, a dropout rate of 0.2 is used for every dropout layer used in the model.

\begin{figure}
    \centering
    \captionsetup{justification=centering}
    \vspace{-3.5ex}
    \includegraphics[width=0.5\textwidth]{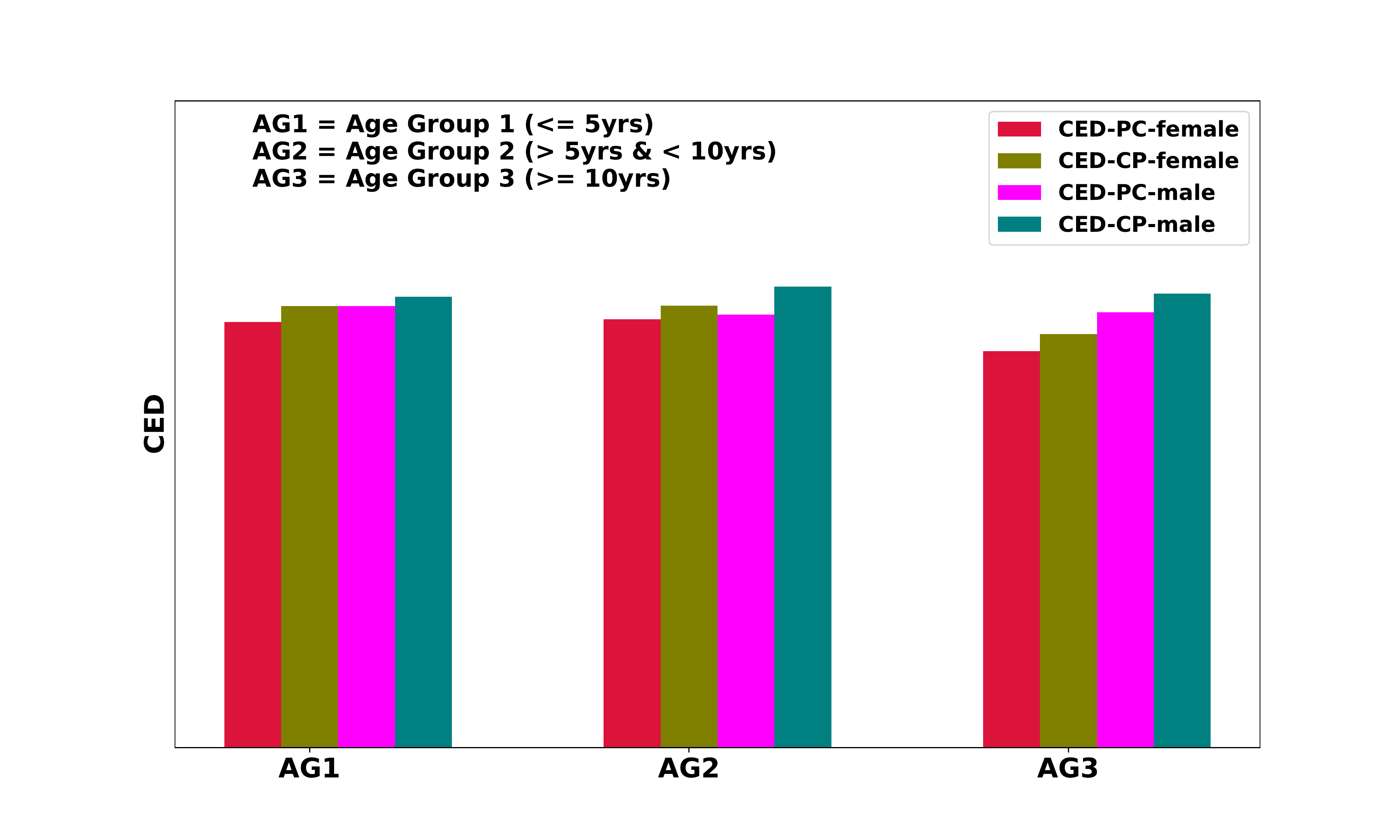}
    \vspace{-4.5ex}
    \caption{Absolute values of CED across age and gender from ADOSMod3}
    \label{fig:barplot}
\vspace{-2.5ex}
\end{figure}

\vspace{-1.5ex}
\subsection{Experimental validation of CED}

We carry out an ad-hoc \textit{real/fake} classification experiment to validate CED as a metric for measuring entrainment. For every \textit{real} sample session we synthesize a \textit{fake} sample session by shuffling the speaker turn while maintaining the dyadic conversation sequence. The hypothesis is more entrainment is expected to be observed in \textit{real} sessions as compared \textit{fake} sessions resulting in the \textit{real} sessions having lesser CED. The classification accuracies are reported in Table 2. The classification experiment steps are as follows:
\begin{itemize}[leftmargin=*,labelindent=1.5mm,labelsep=1.3mm]
    \item We calculate CED measure for every consecutive turn pair for the \textit{real} and \textit{fake} sample session. 
    \item We compare the average CED distance from all the turn pairs for the \textit{real} and \textit{fake} session, the sample sessions are correctly classified if CED of \textit{real} session is lesser than \textit{fake} session.
    \item The experiment is repeated 30 times to eliminate any bias introduced while randomly shuffling the speaker turns.
\end{itemize}

As baselines, we use three distance measures computed between the extracted turn-level pretrained embeddings: smooth L1 distance~\cite{nasir2020modeling} (Baseline 1) and two measures introduced in \cite{lahiri2022interpersonal}, namely, DTWD (Baseline 2), and SCDC (Baseline 3).



\begin{table}[t!]
\begin{center}
\captionsetup{justification=centering}
\caption{Correlation experiment between CED and clinical scores relevant to ASD (bold figures imply statistical significance, $p<0.05$ )\\
\label{tab:correlation}
\textit{(CP:~child to psychologist, PC:~psychologist to child)}}
\resizebox{0.48\textwidth}{!}{
\begin{tabular}{ c|c|c|c|c } 
\hline
\hline
\multirow{3}{*}{Clinical scores} & \multicolumn{4}{c}{Pearson's correlation} \\ 
\cline{2-5}
& \multicolumn{2}{c|}{CED-PC} & \multicolumn{2}{c}{CED-CP} \\ 
\cline{2-5}
& $\rho$ & \textit{p}-value & $\rho$ & \textit{p}-value \\
\hline\hline
VINELAND ABC & -0.061 & 0.237 & 0.012 & 0.827\\
VINELAND Social & -0.021 & 0.345 & 0.071 & 0.073\\
VINELAND Communication  & \textbf{-0.158} & \textbf{0.003} & 0.043 & 0.428\\ 
\hline
CSS & \textbf{0.222} & \textbf{0.004} & 0.023 & 0.672\\
CSS-SA & \textbf{0.231} & \textbf{0.012} & 0.03 & 0.472\\ 
CSS-RRB & 0.158 & 0.055 & 0.091 & 0.262\\
\hline
\hline
\end{tabular}}
\end{center}
\vspace{-5.5ex}
\end{table}

\vspace{-2.5ex}
\subsection{Experimental evaluation}

In this experiment, we calculate the correlation between the proposed CED measure and the clinical scores relevant to ASD in Table \ref{tab:correlation}. Since CED is  directional in nature, we compute the correlation metric in both the directions \textit{child to psychologist} and \textit{psychologist to child}. We report the Pearson's correlation coefficient~($\rho$) and also the corresponding $p$-value, to test the null hypothesis that there exists no linear association between the proposed measure and the clinical scores. Amongst the clinical scores, \textit{VINELAND} scores are designed to measure adaptive behaviour of individuals, while \textit{VINELAND} \textit{ABC} stands for Adaptive Behaviour Composite score, \textit{VINELAND social} and \textit{VINELAND communication} are adaptive behavior scores for specific skills of socialization and communication. \textit{CSS} stands for Calibrated Severity Score which reflects the severity of ASD symptoms in individuals. \textit{CSS-SA} and \textit{CSS-RRB} reflects ASD symptoms severity along 2 domains of \textit{Social Affect} and \textit{Restrictive and Repetitive Behaviours}. The details of the clinical scores related to ASD are described in \cite{bishop2017autism,gotham2009standardizing,hus2014standardizing}.

\par We also report the absolute values of the proposed CED measure~(both directions) for different gender and different age-groups. We partition the dataset across 3 age groups of \textit{Group 1: $\leq 5yrs$, Group 2: $>5yrs$ \& $\leq 10yrs$, Group3: $>10yrs$ } and for each of the age groups we report the directional CED measure for male and female subgroups in Fig.~\ref{fig:barplot}.

\vspace{-2.5ex}
\section{Results and Discussion}
\label{sec:result}
\vspace{-0.5ex}

The results reported in Table \ref{tab:classification} reveal that achieves better performance in identifying \textit{real} and \textit{fake} sessions with respect to the baseline methods in both Fisher and ADOSmod3 corpus in terms of classification accuracy, which validates the use of CED as a proxy metric for measuring entrainment.  

\par Results in Table \ref{tab:correlation} show that VINELAND communication score is negatively correlated with psychologist$\rightarrow$child CED with significant statistic, which stands consistent with the definition of CED, since higher CED signifies lower entrainment. CSS and CSS-SA scores are reported to be positively correlated with CED. It is interesting to note that while psychologist$\rightarrow$child CED  is capturing signals with meaningful interpretations, no such evidence is reported from child$\rightarrow$psychologist CED measures. A possible explanation can be since the model is trained with dyadic conversations from adults in Fisher corpus, the model is unable to capture the nuances of interactions involving children which is reflected in these results. It is also worth mentioning while there exists a significant correlation between CSS, CSS-SA and psychologist$\rightarrow$child CED, CSS-RRB also shows weak evidence of positive correlation with psychologist$\rightarrow$child CED.

\par In Table \ref{fig:barplot}, the distributions for absolute values of CED are reported across gender and age-groups. Both directional CED are always seen to have lesser mean values in females as compared to males, which reiterates the claim reported in \cite{fombonne2020camouflage} that women are better at disguising autism symptoms than men. Across age-groups,the experimental results donot show any discernable observation from CED in both directions in male children, however female psychologist$\rightarrow$child CED is shown to decrease with an increase in age, which also supports the claim presented in \cite{fombonne2020camouflage}.

\par We also investigate the weights of the activations from the cross-encoder attention layer to understand which parts of the speaker turns are emphasized by the attention heads to extract meaningful signals. Attention activation heatmaps from cross-encoder 1 and 2 reported in Fig.~\ref{fig:attention} show attention layers attend to initial few timeframes from the second speaker turn which supports the claim mentioned in \cite{nasir2020modeling} and domain theory that initial and final interpausal units from second and first speaker respectively are a rich source of signals related to entrainment. 

\vspace{-4ex}
\section{Conclusion}
\label{sec:conclusion}
\vspace{-1.5ex}
In this work we introduce a novel context-aware approach (CED) to measure vocal entrainment in dyadic conversations. We use a combination of convolutional neural networks and transformers to capture both short-term and long-term context, and also employ a cross-subject attention module to learn interpersonal entrainment related information from the other subject in a dyadic conversation. We validate the use of CED as a proxy metric for measuring entrainment by conducting a classification experiment to distinguish between \textit{real}~(consistent) and \textit{fake}~(inconsistent) interaction sessions. We also study the association between CED and clinically relevant scores related to ASD symptoms by computing the correlation metric. We also report the mean absolute value of directional CED across gender and different age-groups to understand if the entrainment pattern of the children varies across gender or age-group or not.
In this work, we use a self-supervised embedding for feature extraction, it will be interesting to see if other context-based pre-trained embeddings yeild similar performance in capturing entrainment. We also face difficulties in deploying entrainment embeddings learnt on Fisher for ADOSMod3 dataset and thus we plan to investigate domain-specific entrainment embeddings for understanding behavioral traits.

\bibliographystyle{IEEEbib}
\bibliography{strings}

\end{document}